# The Influence of the Generator's License on Generated Artifacts

Carsten Kolassa and Bernhard Rumpe

Software Engineering
RWTH Aachen University, Germany
http://www.se-rwth.de/

**Abstract.** Open sourcing modelling tools and generators becomes more and more important as open source software as a whole becomes more important. We evaluate the impact open source licenses of code generators have on the intellectual property (IP) of generated artifacts comparing the most common open source licenses by categories found in literature. Restrictively licensed generators do have effects on the IP and therefore on the usability of the artifacts they produce. We then show how this effects can be shaped to the needs of the licensor and the licensee.

## 1 Introduction

Open source has become more and more important in the last years and has been adopted widely in the consumer as well as the industrial market. In 2008, 85% of all enterprises were using open source software [15], more recent studies give numbers up to 98% [31].

Open Source is not only used in business but also in development. According to [16] 76% of all developers have used open source technology for some of their tasks.

The reasons to develop open source software vary greatly. Some companies provide open source software to rapidly grow their user base [18] and thus creating an industry standard. Other reasons are getting a community invested into the project to create an ecosystem of supporting software (e.g. Plugins) arround the original project or to allow external validation of the software (e.g. every user could potentially security audit open source software). Individuals develop open source software to show off their skill, to show their work to the world, for altruistic reasons, or for potential rewards in the future [17]. Academic institutions provide their software under an open source license to gain a user base, and to foster the use of new approaches in industry. Some companies have a dual license approach where they provide an open source version of the software to gain a user base but offer more flexible licenses and customization to business users [29].

The open source licenses reflect this different reasons to open source software and to use it and differ from each other considerably [26].

It is therefore important to chose the right license that is consistent with the reasons you license the software as open source. This is especially true for generators as there are more potential pitfalls than with other software as the generator's license can have an influence on the intellectual property rights of the generated artifacts.

For example a car manufacturer will not use a generator or compiler when the license of the produced code threatens the intellectual property (IP) of his other code or enforces a logo to be placed on the car.

To give an overview about the potential license choices for generators we list the most common licenses and classify them according to categories found in literature. We then show which impact the licenses have on the artifacts if applied to a generator.

Our research questions are:

**RQ1:** Which impact can the open source license choice for the generator have on the license of the artifacts?

**RQ2:** What possible solutions can be used to counteract unwanted effects on the IP of the artifacts?

The paper is structured as follows: Section 2 reviews the related work. In section 3, we describe the licenses and describe their characteristics as found in literature. Section 4 shows which influence those characteristics have on the license/ownership of the generated artifacts if the license is applied to a software generator and how this influence can be shaped to meet the needs of the licensor. Section 5 concludes the paper.

## 2 Related Work

Various publications analyse the rationale behind license choices and give guidelines which license is suitable for which project.

[22] examines the scope of licensing in open source and lists the various considerations that determine the license of open source projects. While [24] gives a guide to choosing an open source license in a commercial context.

[21] examines the licenses and their implications in great detail which negative and positive implication they have on projects in general, the reasons to choose a particular license but aso how to draft an own open source license.

[23] gives an overview about trademark, patent, and copyright law in relation to open source and shows how to choose a commercial or open source license. It examines the implications of linking code covered by the gpl and the implications of creating derivative work but only from a developer perspective as well as from a business perspective, but does not cover model driven development.

[30] shows what motivates businesses to provide open source, examines common open source licenses and how they relate to community and corporate interests. It also classifies open source licenses and gives an overview of the implications of licenses for the IP of the code they cover using examples.

[28] presents how the license choice impacts interest into a project and the development activity.

[12] analyses the open source development paradigm and shows differences and similarities of open source development and licensing to non open source approaches.

However, none of the above papers and books examines the intellectual property situation of artifacts that are generated or created by open source projects as separate case.

## 3 Most Common Licenses

In this section we describe the most common open source licenses and and show their characteristics. We chose the licenses by looking at the black duck license usage statistics [10]. This statistics are calculated from the black duck KnowledgeBase which includes one million open source projects from more than 7500 sites.

We looked at the 10 most widely used licenses and decided to exclude the Artistic License as it is mostly used in the context of the Perl Scripting language and the Microsoft Public License because of its similarity to the Eclipse Public License (EPL) which is more important in the context of model driven development (as many widely used open source software for model driven development use this license) and we decided to look only at the most recent version of each license and exclude older versions which gives us 6 potential licenses for our evaluation. We made the choice to include a 7th license the GNU Affero General Public License as it addresses the privacy loophole which isn't addressed by the other licenses.

### 3.1 Comparison

**Permissions for reuse:** We use the classification in [30] and differentiate between three different types of licenses called Permissive, Weak Copyleft and Strong Copyleft.
- **Permissive:** These licences permit the redistributor to restrict access to the modified source code (make the modified source code closed source) and to put the changed software under a different license (even a proprietary license).
- **Weak Copyleft:** The license of the software cannot be changed but it only applies to the software that is directly derived from the original software e.g. software that incorporates copies of source code from the original software.
- **Strong Copyleft:** Every software that links or otherwise incorporates code from a software licensed under a strong copyleft license needs to be published under a compatible license [27]. Strong copyleft licenses are often called viral as linking to one strong copyleft library forces the whole project to be put under a compatible license.

**Patent license:** Some Open Source licenses automatically include a patent granting clause that grants a non-exclusive, worldwide, royalty-free patent license for all patents a contributor holds and that affect his contribution to the project. In other words if a contributor contributes code to a project that mandates a license grant and the code he contributes would be covered by a patent he holds he needs to give a license to the users of the project without charging them for it.

**Enhanced Attribution:** All open source licenses specify that that the anyone who distributes or modifies the software needs to give credit to the original authors. "Enhanced attribution" means that the license specifies the form of the credits in a way that goes beyond just giving credit like the attribution clause in the original BSD license that specifies that a special acknowledgement needs to be added to all advertising materials mentioning the use of the licensed software or a feature of the software [2]

**Privacy Loophole/Provider Loophole[19]:** If someone modifies an open source software and just uses it or sells its use e.g. a service provider who sells the use of a webclient, there is normally no obligation to make the changes available to the community. If this loophole is closed on the other hand the changes must be made available.

| License | Permissions for reuse | Patent license | Enhanced Attribution | Privacy Loophole |
|---|---|---|---|---|
| Apache License [3] | Permissive | grants | No | No |
| 3-Clause BSD license [9] | Permissive | doesn't grant | No | No |
| MIT License [1] | Permissive | doesn't grant | No | No |
| Eclipse Public License [5] | Weak Copyleft | grants | No | No |
| Lesser GPL (LGPLv3)[8] | Weak Copyleft | grants | No | No |
| Gnu Public License (GPLv3) [7] | Strong Copyleft | grants | No | No |
| Affero General Public License (AGPLv3) [6] | Strong Copyleft | grants | Enhanced[1] | Yes |

Table 1: Comparison of the different licenses.

### 3.2 Dual Licensing/Multi Licensing

Dual licensing is the practice to distribute software under two different licenses [29] while multi licensing is the practice to distribute it under more than two licenses.

Dual Licensing is a common business model in open source. Examples for famous projects with dual licenses are:

- Qt [29]
- MySQL [29]
- Asterisk [11]

---

[1] The license allows to add clauses to require the "preservation of specified reasonable legal notices or author attributions in that material or in the Appropriate Legal Notices displayed by works containing it" [6].

– Sendmail [11]

The software is normally distributed freely under a restrictive open source license that allows the open source community to participate on the development but that makes it difficult to use the software in a commercial environment. But the software is also available under a proprietary software license for a license fee, which allows creating proprietary applications that are based on it (e.g. commercial applications that use the QT library which would't be possible using the open source version) or that allows OEMs (Original Equipment Manufacturers), ISVs (Independent Software Vendors), VARs (Value Added Resellers) to combine and/or distribute the software together with or as commercially licensed software (like in the case of MySQL [25]).

It is also possible to multi license only parts of a software it sometimes makes sense to license a single file under two different open source licenses for example to allow its use in two different projects whose licenses are incompatible to each other or to put some parts of the software under a less restrictive license to allow its reuse while the rest of the software remains under a restrictive license.

## 4 The Ownership of the Generated Code

The question under which license the generated artifacts are is a very important one. The relationship is tricky as the generated artifacts are derived from the models (that are owned by the user of the software) but to some extent also include parts from the generator itself and usually work together with a runtime environment and libraries.

### 4.1 Are Generated Artifacts Derivative Work?

Derivative work according to US copyright law is: "work based upon one or more preexisting works" [4], in software this includes work that contains substantial chunks of source code from another work. Generated source code is therefore derivative work when source code of the original program in our case the generator was used, modified, translated or otherwise changed in any way to create the new program. If the open source license requires that derivative work needs to be published under the same or a compatible license and the generated artifacts are derivative work they automatically need to be open source when they are redistributed, this is the case for all copyleft licenses.

It is important to notice that the mere output of a program itself is not covered by the license of that program. Only if the output constitutes derivative work of the generating program there is this legal dependency between generator output and generator.

Figure 1 shows an example where the generator copies chunks from its own code into the generated artifact which means that the artifact is derivative work.

An example for such a generator is bison the GNU parser generator which copies parts of itself into the parser source code it generates. Therefore every parser generated with bison would be derivative work of bison but the license of bison makes an explicit exception to the GPL [14], that allows to include the bison parser generator's source code in other projects without having to put them under the GPL.

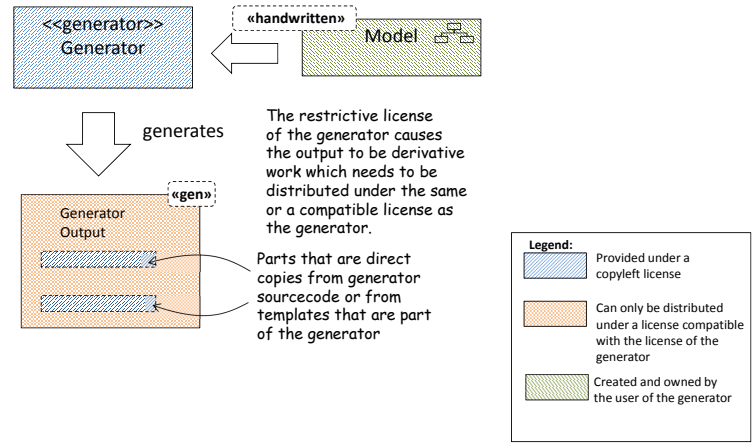

Fig. 1: Relationship of generator and artifacts that causes the artifacts to be derivative work.

Another example for generators that include parts of themselves into generated code are generators based on the Monticore [20] language workbench. Monticore uses templates to generate source code these templates for example generate class definitions and their content is directly put into the generated artifacts. If those templates are licensed under a copy left license the generated code is derivative work and needs to be treated accordingly.

Both weak and strong copyleft licenses impose the restriction that derivative work that includes substantial parts of the original work needs to be put under the same or a compatible license only permissive licenses don't have that restriction.

### 4.2 Dependencies on Libraries

For strong copyleft licenses it isn't enough that the work is not derivative to cause implications of the license to the work depending on it. Strong copyleft licenses define a term "Corresponding Source" which is the code for shared libraries and dynamically linked subprograms that use the work. This "Corresponding Source" needs to be licensed under the same license if the original work is licensed under a strong copyleft license. An example would be a library that is licensed under the GPL and that is used as a shared library by generated artifacts (see Figure 2). These artifacts then need to be licensed under the GPL as well.

For weak copyleft licenses there is no such requirement as merely using a library imposes no effect on the license of the artifacts.

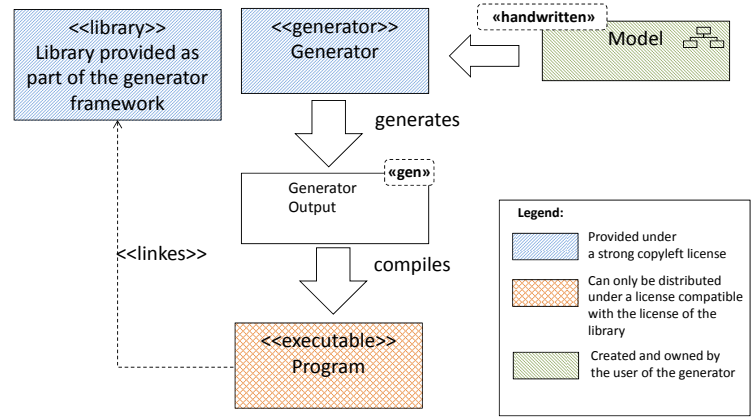

Fig. 2: Relationship of generator and artifacts that causes a dependency between program and library which causes the program to be under the same license as the library it links to.

### 4.3 Explicitly shaping the license implications of open source licenses on generated code.

The license implication both for incorporating source of the generator in generated artifacts as well as having dependencies to strong copyleft licensed code can create practical problems for users of the generator.

For example a company that uses such a generator to create a product would need to license it as open source under a compatible license. This prevents some business models, e.g. the software could not be distributed as closed source. This is often not intended as the licensor sometimes wants to only protect the generator but not the artifacts.

We identify the following ways to explicitly prevent these dependencies from having practical effects:

*Adding an exception in the license.* In this approach en exception is added to a restrictive license to allow exactly the usage of dependent or derivative artifacts that is intended by the licensor. This is the approach that the GNU parser generator used it has the advantage that the intentions of the licensor are clear as it is clearly stated what impact the license of the generator should have on the dependent artifacts.

The disadvantage of this approach is that a modified version of a license (adding an exception is a modification) is a new license. This creates problems for distributors of open source software as they normally use automated approaches to package software which need standardized licenses with clear compatibilities and incompatibilities to other open source licenses.

Users as well as distributors need to check these non standardized exceptions and handle them accordingly this creates additional effort.

*Duallicensing or multilicensing the files the artifacts are derived from or dependent on.* If the generator is under a copyleft license as a whole the licensor can still release parts of it under a less restrictive license. The licensee can then choose the license that applies as both are valid. This solution is recommended by [13] for website templates but can also be applied to generating software. In the case of Monticore based generators for example this means releasing the templates under a permissive license as well while the code parts that don't affect the IP of the generated artifacts are only released under a restrictive license.

*Preventing the dependency from having practical effects by using a permissive license for the files the artifacts are derived from or dependent on.* The third way to prevent dependencies to have practical effects is using a permissive license for the whole project if possible. This has the disadvantage that others can freely use the code of the original work even commercially or in own generators which is often not wanted.

### 4.4 Summary

**RQ1:** Which impact can the open source license choice of the generator have on the license of the artifacts?

**Answer:** The license can have an influence on the artifacts. When the artifacts are derivative work in the case of Weak-Copyleft or Strong-Copyleft licenses or when the artifacts have dependencies on libraries that are part of the generator framework e.g. dependencies on a library in the case of Strong-Copyleft licenses.

**RQ2:** What possible solutions can be used to counteract possibly detrimental effects?

**Answer:** There are three ways to shape minimize the restrictions:
- Adding an exception to the license of the generator.
- Dual licensing the code that creates the dependency under a more permissive license.
- Using a more permissive license that prevents the problem in the first place.

## 5 Conclusion

The license of the generator can have an impact on generated artifacts. Generator providers need to take that into account when choosing the license of their generator. The effects can be positive and deliberate for example to make the open source version of a generator less attractive when the dual licensing business strategy is employed and thus forcing commercial users to license the commercial version of the generator while still being able to benefit from the open source community. But they can also have detrimental effects when a generator that is only available under a restrictive license creates artifacts that are derivative work or have dependencies that cause that the artifacts cannot be used in a commercial environment although the licensor wants to allow that and only intended to protect the generator's code not the artifacts it creates. It is therefore important to know the effects of the license choice on the artifact and to apply the shown solutions if the effects are not as intended.